\documentclass[12pt]{iopart}

\usepackage{graphicx,iopams,bbold}

\usepackage[all]{xy}

\newcommand{\eq}{&=&}

\newcommand{\eqnn}[1]{\begin{eqnarray}#1\end{eqnarray}}

\begin{document}

\title{Optimal Quantum Estimation for Gravitation}

\author{T. G. Downes$^1$, G. J. Milburn$^1$ and C. M. Caves$^2$}

\address{$^1$ Centre for Engineered Quantum Systems, School of Mathematics and Physics, The University of Queensland, St Lucia, QLD 4072, Australia}

\address{$^2$ Center for Quantum Information and Control, University of New Mexico, MSC07-4220, Albuquerque, New Mexico 87131-0001, USA}

\ead{downes@physics.uq.edu.au}

\begin{abstract}
Here we describe the quantum limit to measurement of the classical gravitational field. Specifically, we write down the optimal quantum Cramer-Rao lower bound, for any single parameter describing a metric for spacetime. The standard time-energy and Heisenberg uncertainty relations are shown to be special cases of the uncertainty relation for the spacetime metric. Four key examples are given, describing quantum limited estimation for: acceleration, black holes, gravitational waves and cosmology. We employ the locally covariant formulation of quantum field theory in curved spacetime, which allows for a manifestly spacetime independent derivation. The result is an uncertainty relation applicable to all causal spacetime manifolds. 
\end{abstract}


\maketitle

\section{Introduction}

The geometry of spacetime is determined by physical measurements made with clocks and rulers or, more generally, using quantum fields, sources and detectors. As these are physical systems, the ultimate accuracy achievable is determined by quantum mechanics. In this paper we obtain a parameter-based quantum uncertainty relation, bounding knowledge of the spacetime metric.

In this paper the gravitational field is treated entirely in accordance with classical general relativity. In general relativity, the gravitational field is a manifestation of the geometry of spacetime, which in turn is described by a metric. The metric gives the proper times and proper distances between spacetime events. It is the infinitesimal invariant interval between two spacetime events and is given by \cite{mtw}:
\eqnn{
ds^2\eq g_{\mu\nu}(x)dx^{\mu}dx^{\nu}
\label{metric}
}
Where $g_{\mu\nu}(x)$ is the metric tensor, with indices $\mu,\nu=0,1,2,3$ for the time and three spatial components. The $dx^{\mu}$ are the infinitesimal coordinate distances. We assume the Einstein summation convention, where repeated upper and lower indices are to be summed over. 

We know from Heisenberg that there is an uncertainty relation between position and momentum. It states that the product of the uncertainty in position and the uncertainty in momentum, must always be greater than a constant. In terms of the variances of position and momentum, Heisenberg's uncertainty relation is written as:
\eqnn{
\langle(\Delta \hat{x})^2\rangle\langle(\Delta \hat{p})^2\rangle\geq\frac{\hbar^2}{4}
} 
This relationship bounds knowledge of the position and momentum of a quantum system, where $\hbar$ is the reduced Plank's constant. The Heisenberg uncertainty relation is derived in standard quantum mechanics, where position and momentum are both represented as Hermitian operators. 

A similar relation also exists between time and energy:
\eqnn{
\langle(\delta t)^2\rangle\langle(\Delta \hat{H})^2\rangle\geq\frac{\hbar^2}{4}
\label{teur}
} 
Unlike position, time in standard quantum mechanics is a classical parameter. The time-energy uncertainty relation can be considered an example of quantum parameter estimation. There one tries to estimate a classical parameter, in this case time, by making measurements on a quantum system sensitive to it. In the time-energy uncertainty relation (\ref{teur}), $\langle(\delta t)^2\rangle$ is the classical variance of the estimate of $t$.

The most common way to make quantum mechanics compatible with classical relativity, is to demote position to a parameter, just like time in the previous example. Physical systems can then be thought of as living on, and interacting with, the classical spacetime manifold. In this relativistic context, the Heisenberg uncertainty relation should also be framed in the parameter estimation context \cite{Braunstein:1994}.

This way of thinking was used by Braunstein, Caves and Milburn \cite{Braunstein:1996}, to develop optimal quantum estimation for spacetime displacements, in flat Minkowski spacetime. In spacetime, not only can one move a fixed proper distance or time, one can also boost and rotate. Quantum parameter estimation was thus also developed for the parameters corresponding to these actions \cite{Braunstein:1996}. The results were developed with the quantised electromagnetic field as the measurement system. They show that the estimate of spacetime translations may be made more accurate, if the uncertainty in the number operator is made very large. 

Before we describe quantum parameter estimation in more detail, we ask: can any insight be found by applying the Heisenberg uncertainty relation, in its parameter based form, directly to a proper distance? It was along these lines that Unruh \cite{UNRUH:1986p358} derived an uncertainty relation for the $g_{11}$ component of the metric tensor. The calculation was simple and made the important point that in general relativity the coordinate system is arbitrary and so there should only be quantum uncertainty in the proper time and the proper distance. As these are in turn related to the metric via (\ref{metric}), any uncertainty in the proper distance is equivalent to uncertainty in the metric. By applying the Heisenberg uncertainty relation to a proper distance, a simple but insightful uncertainty relation was found for one component of the metric. We write down the Unruh uncertainty relation for the $xx$ component of the metric, in terms of variances as:
\eqnn{
\langle(\delta g_{11})^2\rangle\langle(\Delta\hat{T}^{11})^2\rangle\geq\frac{\hbar^2}{V^2}
\label{umur}
} 
The key finding of this uncertainty relation is the inverse proportionality to $V^2$, the square of the four-volume of the measurement. The conjugate variable to $g_{11}$ is the corresponding component of the quantised stress-energy tensor $\hat{T}^{11}$, in this case the pressure in the $x$-direction. 

In this paper we consider a more general context for deriving an uncertainty relation for the metric, by formulating it as a problem in quantum estimation theory. The metric $g_{\mu\nu}(x)$ is defined for each point $x$ on the manifold. If the quantum measurement system occupies some four-volume $V$, then the system will depend on the the metric at every point in that region. If we consider the metric to be an arbitrary function on the manifold, then we need to estimate an infinite number of parameters to define it completely. Instead we will consider regions of spacetime, which can be described by metrics, defined by a finite number of parameters $\theta_1,\theta_2\dots\theta_N$. For example the Schwarzschild metric, which describes the spacetime around a static non-rotating black hole, is defined by only one parameter $M$, the mass of the black hole. The task is to estimate the parameters by making measurements on physical systems living on the spacetime manifold. After we derive the uncertainty relation in this parameter based context, we can consider the limit of measuring the gravitational field at a singe point in space-time. In this case the parameters simply become the individual components $g_{\mu\nu}$ of the metric (\ref{metric}).  

\begin{figure}[h]
   \centering
   \includegraphics[height=3.8cm]{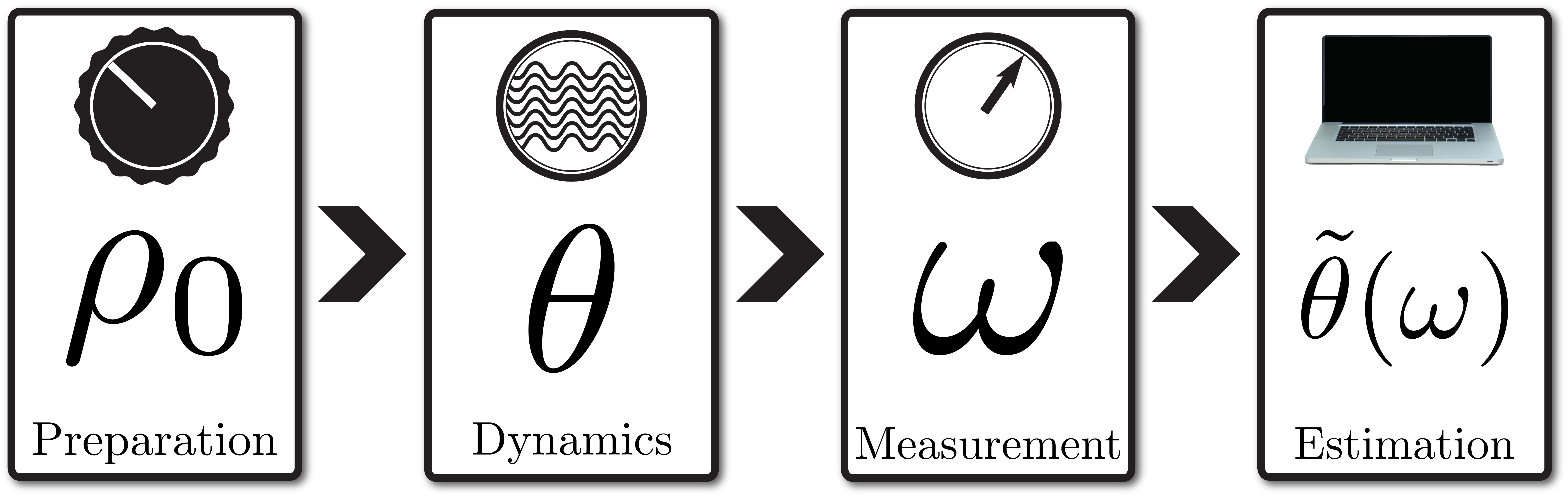} 
   \caption{The scheme for quantum parameter estimation.}
   \label{qpe-scheme}
\end{figure}

In this paper we shall consider estimating an individual parameter of a spacetime metric and so we shall only consider single parameter quantum estimation. The general schema for quantum parameter estimation is represented in Figure \ref{qpe-scheme}.
An initial quantum state, represented by a density operator $\rho_0$, is evolved by a unitary transformation $\hat{U}(\theta)$ dependent on the parameter $\theta$ of interest. Measurements are then made on the system, producing measurement results $\omega$, which are in turn fed into an estimator $\tilde{\theta}(\omega)$ of the parameter. 

We consider generalised measurements, described by positive operator valued measures. These are Hermitian operators with the property:
\eqnn{
\int \hat{E}(\omega)d\omega=\mathbb{1}
}

Where $\omega$ is the measurement result and $\mathbb{1}$ is the identity operator. The outcomes of a particular measurement follow a probability distribution $p(\omega |\Theta)$ conditional on the actual value of the parameter $\Theta$. The probability distribution for the outcomes $\omega$ can be calculated as:
\eqnn{
p(\omega|\Theta)\eq\mathrm{Tr}\left(\hat{\rho}(\Theta)\hat{E}(\omega)\right)
}
Where $\hat{\rho}(\Theta)=\hat{U}(\Theta)\rho_0\hat{U}^{\dagger}(\Theta)$ is the state after the $\theta$ dependent interaction. 

The problem of estimating the parameter $\theta$ is essentially that of choosing a value $\tilde{\theta}$ to fit the probability distribution $p(\omega|\tilde{\theta})$ to the actual measured values. A common example is the maximum likelihood estimator, which is the choice of $\tilde{\theta}$ which retrospectively maximises the probability, of the observed measurement values. 

The variance of any estimate of the parameter $\theta$, based on the distribution of the observed measurements, is bounded below by the classical Cramer-Rao lower-bound \cite{Braunstein:1996}: 
\eqnn{
\langle(\delta \tilde{\theta})^2\rangle\geq\frac{1}{F(\Theta)}
}
where $F(\Theta)$ is the Fisher information for the measurement, given by:
\eqnn{
F(\Theta)=\int d\omega\frac{1}{p(\omega|\Theta)}\left(\frac{\partial p(\omega |\theta)}{\partial \theta}\bigg|_{\theta=\Theta}\right)^2
}
The key ingredient is the rate of change of the probability distribution with respect to the parameter. A larger response due to a change in the parameter naturally gives rise to better estimation of the parameter. The rate of change of the probability distribution can be calculated as:
\eqnn{
\frac{\partial p(\omega |\theta)}{\partial \theta}\eq\mathrm{Tr}\left(i[\hat{\rho}(\theta),\hat{h}]\hat{E}(\omega)\right)
}
Here $\hat{h}$ is the generator of the unitary transformation $\hat{U}(\theta)=e^{i\hat{h}\theta}$. 

We have described quantum parameter estimation in the Schr\"odinger picture where the density operator depends on the parameter. Due to cyclicity of the trace used to calculate both the probability distribution and its rate of change, the Heisenberg picture can equally be used. The key component is the generator $\hat{h}$ which generates changes in the probability distributions due to changes in the parameter. This can be seen directly for pure states, when one optimises over all possible measurements. The optimal quantum Fisher information is given by:
\eqnn{
F(\Theta)=\frac{4\langle(\Delta\hat{h}[\Theta])^2\rangle}{\hbar^2}
} 
Where $\langle(\Delta\hat{h}[\Theta])^2\rangle$ is the quantum variance of the generator $\hat{h}$ taken in the initial state. The optimal quantum Cramer-Rao lower bound can then be expressed as:
\eqnn{
\langle(\delta \tilde{\theta})^2\rangle\langle(\Delta\hat{h}[\Theta])^2\rangle\geq\frac{\hbar^2}{4}
}
One can now easily construct examples by simply identifying parameters and their corresponding generators. For example, the Hamiltonian is the generator of time translations, which gives the time-energy uncertainty relation (\ref{teur}) presented in the introduction. Another important example is the number operator and phase which is the basis of Heisenberg limited phase estimation.

\section{An Uncertainty Relation for the metric $g_{\mu\nu}$}

The physical systems we consider here are quantum fields, for example the Dirac field for electrons and the electromagnetic field for photons. For clarity we present our derivation using the free scalar field, the generalisation to interacting and higher spin fields should follow a similar argument. 

We consider only measurements on spatio-temporal scales large enough, and energy scales small enough, such that the quantisation of the gravitational field itself can be safely ignored. Matter and non-gravitational energy are treaded with quantum mechanics whereas gravity is treated using classical general relativity. 

We also work in the test-field approximation, where the gravitational field of the probe is ignored completely. For example one might choose to ignore the gravitational field of the laser in a gravitational wave interferometer. It is important to note however, that the action of the probe fields on the spacetime manifold will only strengthen the bound we derive below. That is, if a high energy field were used to probe the structure of spacetime, then the back-reaction of the field on the spacetime, would further limit the accuracy of the measurements, preventing our inequalities from being saturated. The problem of back-reaction when measuring the structure of spacetime has already been studied in some detail \cite{ngnvandam}. 

Apart from the test-field approximation, the quantum theory for the fields we consider here takes classical general relativity fully into account. We employ the locally covariant formulation of quantum field theory on curved spacetime as presented by Brunetti, Fredenhagen and Verch \cite{bfv}. Local covariance refers to the global property of having the same physics under all coordinate systems, i.e.~general covariance, together with the local property that if two localised spacetime regions are equivalent, then they should describe the same physics regardless of what's happening at distant regions of the universe. In this sense the locally covariant formulation fully takes into account the locality and covariance of general relativity. We only omit causality-violating spacetimes, which removes certain causal pathologies associated with closed time-like curves and time machines. Our results will hold for all causal spacetime manifolds. 

The locally covariant approach was instrumental in completing the perturbative construction of interacting quantum field theory in curved spacetime \cite{brunetti:2000,hollands:2001,hollands:2002}. In a similar way it is now being applied to the problem of perturbative quantum gravity  \cite{fredenhagen:2011}. This is the weak field limit where the quantum part of gravity is a perturbation of a large classical background. The locally covariant approach attempts to solve the problems associated with traditional attempts at perturbative quantum gravity including background independence and renomalisability. Indeed it is being advocated as a strong third approach to quantum gravity alongside string theory and loop quantum gravity  \cite{fredenhagen:2011}. The power of this new formulation is summarised well by Christofer Fewster \cite{fewster:2011}:
\begin{verse}
``\dots we stress that the BFV approach is not simply a matter of formalism: it suggests and facilitates new calculations that can lead to concrete physical predictions\dots" 
\end{verse}
In this paper we present a new example of such a calculation which indeed leads to concrete physical predictions.

A key result of the locally covariant approach is a method for calculating how quantum observables respond to changes in the metric. Say we believe some particular region of the universe to be well described by a metric $g_{\mu\nu}(\theta)$ depending on $N$ parameters $\theta=(\theta_1,\theta_2,\dots,\theta_N)$. For this case the locally covariant approach can be used to calculate how any observable $\hat{E}(\theta)$ will respond to a change in any one of the parameters. The response is evaluated as the rate of change of the observable with respect to the parameter. As we noted in the Introduction, this is just what we need to calculate the quantum Cramer-Rao lower bound.  

\begin{figure}[h]
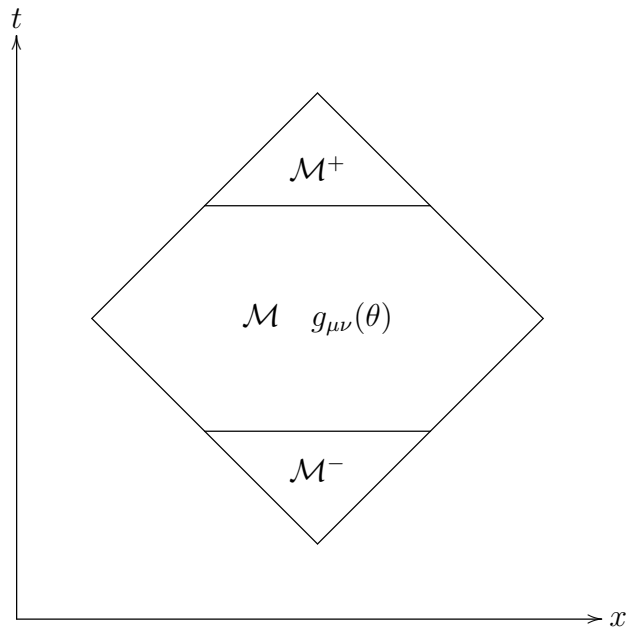

   \centerline{\xy (0,0) ; (0,80)*+{t} **\dir{-} ?>* \dir{>}, 
   	(0,0) ; (80,0)*+{x} **\dir{-} ?>* \dir{>},
	(40,60)*{\mathcal{M}^+},
	(40,20)*{\mathcal{M}^-},
	(40,40)*{\mathcal{M}\quad g_{\mu\nu}(\theta)},
	(10,40) ; (40,70) **\dir{-}, 
	(70,40) ; (40,70) **\dir{-}, 
	(25,25) ; (55,25) **\dir{-}, 
	(25,55) ; (55,55) **\dir{-}, 
	(10,40) ; (40,10) **\dir{-}, 
	(70,40) ; (40,10) **\dir{-} \endxy}
   
   \caption{Spacetime diagram. The middle region is described by a metric $g_{\mu\nu}$ dependant on the parameter $\theta$.}
   \label{spacetime}
\end{figure}

We first consider an arbitrary region of spacetime and divide it into three sub-regions as shown in Figure \ref{spacetime}. The lower region labeled $\mathcal{M}^-$ is the region where the field is prepared. It is in the causal past of the upper region labeled $\mathcal{M}^+$ where measurements are performed. The intermediate region $\mathcal{M}$ is inside the intersection of the causal past of $\mathcal{M}^+$ and the causal future of $\mathcal{M}^-$. 

We consider a diffeomorphism $\phi_s$ which smoothly deforms the metric inside the region $\mathcal{M}$ and acts like the identity everywhere else. The effect of this change on any observable $\hat{E}(\theta)$ in the measurement region can be calculated by the action of a map, known as the relative Cauchy evolution \cite{bfv}. First we construct the observable $\hat{E}(\theta)$ for a particular value of the parameter, $\theta=\Theta$, and then use the relative Cauchy evolution to calculate the rate of change with respect to $\theta$. Specifically the diffeomorphism chosen acts on the metric in the region of interest as: 
\eqnn{
\phi_s g_{\mu\nu}(\Theta)\eq g_{\mu\nu}(\Theta+s)
}
The number $\Theta$ can be thought of as the actual or expected value of the parameter. If we have an unbiased estimator $\tilde{\theta}$ of the parameter $\theta$ then the expected value of the estimate $\langle\tilde{\theta}\rangle=\Theta$.

We construct the measurement operators out of polynomials of localised field operators. For a particular measurement $\hat{E}(\omega|\theta)$, conditional on a parameter $\theta$ in the metric, we can construct the probability distribution of the outcomes $\omega$ as:
\eqnn{
p(\omega|\theta)\eq\mathrm{Tr}\left(\hat{\rho}\hat{E}(\omega|\theta)\right)
\label{pdf}
}
where $\hat{\rho}$ is the density operator representing the state of the system. The relative Cauchy evolution gives the rate of change of the probability distribution with respect to the parameter as \cite{bfv,downesThesis}:
\eqnn{
\frac{d p(\omega |\theta)}{d\theta}\bigg|_{\theta=\Theta}\eq\mathrm{Tr}\left(i\hat{\rho}[\hat{E}(\omega|\Theta),\hat{P}(\Theta)]\right)
\label{dpdtheta}
}
where the operator $\hat{P}(\theta)$ is given by:
\eqnn{
\hat{P}(\theta)\eq\frac{1}{2c}\int_{\mathcal{M}}d\mu_g\hat{T}^{\mu\nu}\frac{dg_{\mu\nu}(\theta)}{d \theta}
\label{bigp}
}
Here $c$ is the speed of light, $\mathcal{M}$ is the region of interest and $d\mu_g$ is the volume-form induced by $g_{\mu\nu}(\Theta)$. $\hat{T}^{\mu\nu}$ is the renormalised stress-energy tensor of the probe field. We identify the operator $\hat{P}(\theta)$ as the infinitesimal generator of changes in the probability distribution due to changes in the parameter $\theta$. The cyclicality of the trace and the simple commutator form of the expression in relation (\ref{dpdtheta}) means that the dependance can either be thought of as in the observable or the state. The standard techniques of quantum parameter estimation, as described in the introduction, can be immediately applied. For pure states, optimisation over all possible measurements, yields the following uncertainty relation:
\eqnn{
\langle(\delta \tilde{\theta})^2\rangle\langle(\Delta \hat{P}[\Theta])^2\rangle\geq\frac{\hbar^2}{4}
\label{mur}
}
This relation gives the optimal quantum limit to measuring any parameter of a spacetime metric. It is locally and covariantly defined and applies to all causal spacetime manifolds. We see that the uncertainty relation (\ref{mur}) is independent of any coordinates used to evaluate it, as the four-volume integral of a scalar quantity, such as (\ref{bigp}), is invariant under coordinate transformations. However the uncertainty relation (\ref{mur}) does depend on the choice of the measurement region $\mathcal{M}$, as well as the behaviour of the stress-energy tensor within that region. To demonstrate the ease with which it can be applied, we shall present a broad range of examples.

\section{Metric Uncertainty Relation Simplified}
Consider now making measurements in a region of spacetime where the metric can be approximated as constant. The only parameters to estimate are then the components of the metric. If we choose to estimate an arbitrary component of the metric, i.e. $\theta=g_{\mu\nu}$ for some fixed $\mu,\nu$, the uncertainty relation (\ref{mur}) becomes:
\eqnn{
\langle(\delta g_{\mu\nu})^2\rangle\langle(\Delta \int_{\mathcal{M}}d\mu_g\hat{T}^{\mu\nu})^2\rangle\geq\hbar^2c^2
\label{psmur}
}
Where in this case there is no sum over the repeated indices as they are to be interpreted as fixed values. We further simplify the uncertainty relation (\ref{psmur}) by assuming the stress-energy tensor has constant variance over the measurement region. In this case one is left with a particularly simple form for the metric uncertainty relation given by:
\eqnn{
\langle(\delta g_{\mu\nu})^2\rangle\langle(\Delta\hat{T}^{\mu\nu})^2\rangle\geq\frac{\hbar^2}{V^2}
\label{smur}
}
where $V=\frac{1}{c}\int_{\mathcal{M}}d\mu_g$ is the four-volume of the measurement with units of $\mathrm{sec}\cdot\mathrm{m}^3$. This form of the fundamental dependance confirms the earlier relation (\ref{umur}) found by Unruh. Indeed the same restrictions were needed to derive (\ref{umur}) as were used to produce (\ref{smur}). It should be noted that the stress-energy tensor $\hat{T}^{\mu\nu}$ in (\ref{smur}) is for the probe field. It is not the stress-energy tensor for the matter distribution which gives $g_{\mu\nu}$ via Einstein's field equations. This is to be expected as (\ref{smur}) is essentially an uncertainty relation for the field, however, as we are estimating the metric with field measurements, uncertainty in the field in (\ref{smur}) has been replaced by uncertainty in the metric.

\section{Estimation of Proper Time and Proper Distance}
We shall now look at a specific case of the relation (\ref{psmur}) for the time-time component of the metric. In this case relation (\ref{psmur}) simply becomes:
\eqnn{
\langle(\delta g_{00})^2\rangle\langle(\Delta \int_{\mathcal{M}}d\mu_g\hat{T}^{00})^2\rangle\geq\hbar^2c^2
\label{tmur}
}
The integral of the energy density over the spatial component of the four volume is the Hamiltonian\footnote{Note that $\hat{T}^{00}$ has units of energy density divided by $c^2$.}:
\eqnn{
\int_{\mathcal{M}}d\mu_g\hat{T}^{00}=\frac{1}{c^2}\sqrt{\langle g_{00}\rangle}\int dt\hat{H}(t)
} 
and so the uncertainty relation (\ref{tmur}) becomes:
\eqnn{
\langle(\delta g_{00})^2\rangle\langle(\Delta \int dt\hat{H}(t))^2\rangle\geq\frac{\hbar^2c^4}{\langle g_{00}\rangle}
\label{tmur2}
}
Now consider the proper time of an observer stationary in the local coordinate system we are working in. The proper time in this case is given by:
\eqnn{
\tau=t\sqrt{g_{00}}
}   
The coordinates have been chosen, so there is no uncertainty in the coordinate time. Using the delta method, the uncertainty in the metric is then related to the uncertainty in the proper time by:
\eqnn{
\langle(\delta g_{00})^2\rangle=4\langle g_{00}\rangle\langle(\delta \tau)^2\rangle/t^2
} 
Inserting this relation into equation (\ref{tmur2}) gives the time energy uncertainty relation generalised to curved spacetime in a local inertial frame:
\eqnn{
\langle(\delta \tau)^2\rangle\langle(\Delta \frac{1}{t}\int dt\hat{H}(t))^2\rangle\geq\frac{\hbar^2c^4}{4\langle g_{00}\rangle^2}
\label{tmur3}
} 
If we assume a time independent Hamiltonian and specialise to flat spacetime where $\langle g_{00}\rangle=c^2$ then the uncertainty relation (\ref{tmur3}) reduces to:
\eqnn{
\langle(\delta \tau)^2\rangle\langle(\Delta \hat{H})^2\rangle\geq\frac{\hbar^2}{4}
} 
This demonstrates that the standard time energy uncertainty relation is a special case of the metric uncertainty relation (\ref{mur}). Using a similar argument one can also derive an uncertainty relation for proper distance $X$.
\eqnn{
\langle(\delta X)^2\rangle\langle(\Delta \hat{P})^2\rangle\geq\frac{\hbar^2}{4}
} 
This is the parametric version of the Heisenberg uncertainty relation where $\hat{P}$ is the momentum in the direction of the displacement.

\section{Estimation of Proper Acceleration}
We begin our examples with a uniformly accelerated observer in flat spacetime. Here it is a property of the observer we are interested in, not the spacetime itself and so the observer effectively fixes the coordinate system. This example is perhaps best interpreted as a constant acceleration drive spacecraft, trying to determine it's own acceleration by local measurements. The natural coordinate system, for a uniformly accelerating observer, is know as the Fermi-Walker transported orthonormal tetrad. It is the coordinate system carried by the accelerating observer, in which he is instantaneously at rest. This coordinate system has limits, it can only be extended at most a distance $c^2/a$ from the observer, where $a$ is the proper acceleration and $c$ is the speed of light. The flat Minkowski metric in these coordinates can be written as \cite{mtw}:
\eqnn{
ds^2\eq -\frac{1}{c^2}(c^2+a\xi^1)^2(d\xi^0)^2+(d\xi^1)^2+(d\xi^2)^2+(d\xi^3)^2
}
Here we take the region of interest to be a 4-cube, centred at the origin, in these coordinates with sides of length $L_{\xi^1},L_{\xi^2},L_{\xi^3}$ and duration $L_{\xi^0}$. To keep the examples simple, throughout the rest of this paper we shall restrict to states of the field where both $\langle\hat{T}(x)\rangle$ and $\langle\hat{T}(x)\hat{T}(x^{\prime})\rangle$ are effectively constant, in the chosen coordinates over the region of interest. With these considerations and choosing $\theta=a$ in (\ref{mur}) the uncertainty relation for the proper acceleration simply becomes:
\eqnn{
\langle(\delta \tilde{a})^2\rangle\langle(\Delta \hat{T}^{00})^2\rangle\geq\left(\frac{3\hbar c^2}{aL_{\xi^0}L_{\xi^1}^3L_{\xi^2}L_{\xi^3}}\right)^2
\label{paur}
}
In agreement with the simplified relation (\ref{smur}) we see that better estimation is achieved by making the apparatus larger. Here in particular one receives an inverse cubed reduction in the minimum uncertainty by extending the apparatus out in the direction of acceleration, in comparison to the strait inverse reduction in the orthogonal directions. We also see that the uncertainty will decrease with increasing acceleration, however if we make the apparatus as large as possible, by inserting the constraint $L_{\xi^1}\leq 2c^2/a$ we find:
\eqnn{
\langle(\delta \tilde{a})^2\rangle\langle(\Delta \hat{T}^{00})^2\rangle\geq\left(\frac{3\hbar  a^2}{8c^4L_{\xi^0}L_{\xi^2}L_{\xi^3}}\right)^2
}
Now the inverse is true, if one has reached an acceleration, such that the size of the apparatus has been limited, then larger accelerations will produce worse estimation. This can be seen in the first instance, from the much stronger dependence on size as compared to the acceleration. 

We see from this example that the optimal procedure would be to extend the accelerometer along the full length of the spacecraft. Measured uncertainty could be compared with the proper acceleration uncertainty relation to test how close the accelerometer was to achieving optimal performance.

\section{Estimating the Mass of a Black Hole}
Consider the problem of measuring the gravitational field, outside a spherically symmetric, non-rotating, massive body. The metric for the empty space outside such an object, is given by the Schwarzschild solution to Einstein's field equations \cite{mtw}. In Schwarzschild coordinates the metric becomes:
\eqnn{
ds^2=-\left(1-\frac{r_s}{r}\right)c^2dt^2+\left(1-\frac{r_s}{r}\right)^{-1}dr^2+r^2d\theta^2+r^2\sin^2\theta d\phi^2
\label{bhm}
}
Here $r_s=2MG/c^2$ is the Schwarzschild radius, where $M$ is the mass as observed from infinity and $G$ is the gravitational constant. These coordinates have the intuitive features that; surfaces of constant $t$ and $r$ have area given by $4\pi r^2$ and the metric (\ref{bhm}) becomes the metric of an inertial observer in flat spacetime, as $r$ becomes large. We choose the region of interest $\mathcal{M}$ to have duration $L_t=t_2-t_1$, length $L_r=r_2-r_1$ and solid angle $L_\Omega=\int_{\theta_1}^{\theta_2}\int_{\phi_1}^{\phi_2}\sin\theta d\theta d\phi$ in Schwarzschild coordinates. 

The uncertainty relation (\ref{mur}) is defined in terms of a four-volume integral of a scalar quantity. Due to this, it is invariant under general coordinate transformations. We shall take full advantage of this invariance and change to more well behaved coordinates, in order to evaluate the uncertainty relation. We transform to outgoing Eddington-Finkelstein coordinates, which are appropriate for describing photons propagating out of the gravitational potential \cite{mtw}. The outgoing null geodesics are labeled with a new coordinate $u=t-r^*/c$ where:
\eqnn{
r^*=r+r_s\ln\left|\frac{r}{r_s}-1\right|
}
In the coordinates $(u, r, \theta,\phi)$ the metric takes the outgoing Eddington-Finkelstein form \cite{mtw}: 
\eqnn{
ds^2=-\left(1-\frac{r_s}{r}\right)c^2du^2-2cdudr+r^2(d\theta^2+\sin^2\theta d\phi^2)
}
The behaviour of the stress-energy tensor over the region of interest, will affect the form of the uncertainty relation after the integral in (\ref{bigp}) has been evaluated. For simplicity in this particular example we restrict to states, such that the two moments $\langle \hat{T}^{00}(x)\rangle$ and $\langle \hat{T}^{00}(x)\hat{T}^{00}(x^{\prime})\rangle$ of the stress-energy tensor, are constant over the measurement region, with respect to the Eddington-Finkelstein coordinates. With these considerations and choosing $\theta=M$ in (\ref{mur}) the uncertainty relation for the mass of the black hole becomes:
\eqnn{
\langle(\delta \tilde{M})^2\rangle\langle(\Delta\hat{T}^{00})^2\rangle\geq\left(\frac{\hbar}{GL_tL_\Omega [r_2^2-r_1^2]} \right)^2
} 
Here $\hat{T^{00}}$ is the energy density divided by $c^2$ in Eddington-Finkelstein coordinates. It is important to note that $L_t, L_{\Omega}, r_1, r_2$ are just a set of numbers which describe the measurement region. They have a simple interpretation in terms of the coordinates used to define them but the same set of numbers could also be used to describe the same region in a different coordinate system. In an alternative coordinate system the description of the region in terms of these numbers would be more complicated, however once one has performed the volume integral in (\ref{bigp}) the same uncertainty relation will result, independent of the coordinate system used. It must be noted that although we used a particular coordinate system to define our measurement region, the choice of a particular measurement region is independent of the coordinates used to describe it. 

In this example we once again see the uncertainty will decrease as the size of the measurement region is increased. The dependance on the distance from the mass is clearly stronger, than that of time or solid angle. This is due to the fact that the gravitation field varies in this direction. In analogy to the previous example, one achieves better estimation by extending the apparatus out in the direction of the massive object.

\section{Quantum Limited Gravitational Wave Detection}
We now consider estimating the amplitude of a gravitational wave. In the linear approximation of Einstein's equations, the metric can be written as the flat Minkowski metric plus a small perturbation.
\eqnn{
g_{\mu\nu}\eq\eta_{\mu\nu}+h_{\mu\nu}
} 
The simplest solutions to the linearised Einstein equations are plane-waves. We write down these wave solutions, in the transverse traceless gauge, as \cite{mtw}:
\eqnn{
h_{\mu\nu}\eq\mathbb{R}\left[A_{\mu\nu}\exp(k_ax^a)\right]
\label{gwm}
}  
The amplitude of the wave $A_{\mu\nu}$ has only two independent components corresponding to the two polarisation states. For our example here, we shall consider a gravitational wave propagating in the $z$ direction and linearly polarised along the diagonal directions in the $x$-$y$ plane. For this case the only non-zero components of the metric (\ref{gwm}) are \cite{mtw}:
\eqnn{
h_{12}=h_{21}=A_{\times}\mathrm{sin}[\omega(t-z/c)]
}
As the wave is a small perturbation $A_{\times}\ll1$ we approximate the four-volume element, as being the volume element of the flat metric. For our measurement region we take a 4-cube centred at the origin, in Minkowski coordinates, with length $L_x,L_y,L_z$ and duration $L_t$. Once again we restrict to states where the first two moments of the stress-energy tensor are constant. When estimating the amplitude of the wave, the uncertainty relation (\ref{mur}) becomes: 
\eqnn{
\langle(\delta A_{\times})^2\rangle\langle(\Delta\hat{T}^{12})^2\rangle\geq\frac{\hbar^2\omega^4}{64c^2L_x^2L_y^2\sin^2[\frac{\omega}{2c} L_z]\sin^2[\frac{\omega}{2} L_t]}
\label{gwur1}
}
If one considers time durations and lengths such that $\frac{\omega}{2}L_t,\, \frac{\omega}{2c} L_z\ll1$, then the sinusoidal functions can be approximated as linear, which simplifies (\ref{gwur1}) to:
\eqnn{
\langle(\delta A_{\times})^2\rangle\langle(\Delta\hat{T}^{12})^2\rangle\geq\left(\frac{\hbar}{2L_xL_yL_zL_t}\right)^2
\label{gwur2}
}
For example the Earth-Sun system will produce gravitational radiation with an average $\omega = 2\times10^{-7}\mathrm{sec}^{-1}$. This allows $L_t$ to be months long and $L_z$ to be close to a light-year in length before the linear approximation to the sinusoidal functions breaks down.

A limitation to the uncertainty relation (\ref{mur}) in dynamical situations like this one, is that it is based on a single measurement and hence misses the averaging that would occur over a continuous measurement. Due to this the uncertainty in (\ref{gwur1}) diverges whenever the transverse length of the apparatus is a multiple of the wavelength. Consider a ring of atoms sitting in the plane perpendicular to the direction of propagation of the wave. Once a complete wavelength has passed they will distort and then return to their original position. If a measurement is only made at this point then no effect will be observed, and hence no information on the amplitude will have been received. Generalising the uncertainty relation (\ref{mur}) to take into account continuous measurements is then the logical next step. It will require analysing the Fisher information matrix and will result in a spectral uncertainty relation \cite{caves}.

\section{Estimating the expansion of the universe}

For our final example we shall consider the spatially flat Friedmann-Robertson-Walker cosmology. This is a universe filled with a uniform density of galaxies. At any instant in time, in the co-moving frame of the galaxies, the universe looks the same everywhere (homogeneous) and in all directions (isotropic). The metric for this universe is given by \cite{mtw};
\eqnn{
ds^2\eq -dt^2+a^2(t)\left[dx^2+dy^2+dz^2\right]
} 
where $t$ is the proper time of an observer co-moving with any of the galaxies. The spatial coordinates $x,y,z$ describe the homogeneous and isotropic surfaces of constant proper time $t$. The function $a(t)$, known as the expansion parameter, is the ratio of the proper distance between any two galaxies at the initial time $t=0$ and the time $t$. For convenience, in this example we work in units such that the speed of light $c=1$.  

During an infinitesimal duration of proper time $dt$ a photon will travel the distance $d\eta=dt/a(t)$. It is convenient to use $\eta$ as the time parameter. We shall consider a universe dominated by matter, in which case the metric becomes;
\eqnn{
ds^2\eq \frac{a_{\mathrm{max}}^2}{4}(1-\cos\eta)^2\left[-d\eta^2+dx^2+dy^2+dz^2\right]
} 
where $\eta$ runs between $0$ at the beginning of expansion to $2\pi$ at the end of recontraction. We wish to estimate the parameter $a_{\mathrm{max}}$ which controls the maximum size the universe reaches before recontraction commences. The uncertainty relation (\ref{mur}) for this parameter becomes:
\eqnn{
\sum_{\mu}\langle(\delta\tilde{a}_{\mathrm{max}})^2\rangle\langle(\Delta\hat{T}^{\mu\mu})^2\rangle\geq\left(\frac{16\hbar}{L_xL_yL_za_{\mathrm{max}}^5\int_{\eta_1}^{\eta_2}d\eta(1-\cos\eta)^6}\right)^2
} 
The integral in the denominator can be performed analytically but perhaps becomes somewhat clearer if expressed in terms of the density of mass-energy $\rho$:
\eqnn{
\sum_{\mu}\langle(\delta\tilde{a}_{\mathrm{max}})^2\rangle\langle(\Delta\hat{T}^{\mu\mu})^2\rangle\geq\left(\frac{4\hbar}{3\pi L_xL_yL_za_{\mathrm{max}}\int_{\eta_1}^{\eta_2}d\eta\rho^{-2}(\eta)}\right)^2
} 
Again better estimation is achieved by increasing the size of the apparatus, as well as for larger values of $a_{\mathrm{max}}$. Better estimation is also made by running the experiment for longer time $\eta$. However for fixed duration, better estimation is made during the period of low energy-density, which occurs at the point of maximum expansion, that is during the period where $a_{\mathrm{max}}$ is actually reached.

\section{Conclusion}

In this paper we have presented the optimal quantum Cramer-Rao lower bound for parameters describing a metric for spacetime. Our specific derivation applies for pure states of the scalar field on an arbitrary causal spacetime manifold. We give four important examples covering the full gamut of relativistic phenomena.  We described quantum estimation for; the acceleration of a uniformly accelerating observer, the mass of a black hole, the amplitude of a gravitational wave and the expansion parameter in a cosmological model. In all these examples the fundamental dependance can be seen as the inverse proportionality to the four-volume of the measurement. This dependance can be seen explicitly when one makes the measurement region sufficiently small, in agreement with earlier work on the subject  \cite{UNRUH:1986p358}. 

The methods developed here can easily be applied to many situations involving the measurement of gravity. The uncertainty relation (\ref{mur}) can be used to benchmark the optimality of experimental proposals involving high precision measurements of gravitational phenomena. By evaluating the form of the uncertainty relation (\ref{mur}) for particular scenarios, one can ascertain which parameters have the biggest impact on the measurement results. Hence one can better understand the tradeoff between cost and impact when attempting to optimise experimental design.

\ack

T.G.D. would like to thank Klaus Fredenhagen and support from DESY, as well as Rainer Verch and support from the University of Lipzig. We also thank Achim Kempf and Tim Ralph for useful discussions.

\end{document}